\documentclass[12pt]{article}

\usepackage{cite}
\usepackage{amsmath,amssymb,amsfonts}
\usepackage{algorithmic}
\usepackage{graphicx}
\usepackage{textcomp}
\usepackage{xcolor}
\usepackage{authblk}

\title{Bayesian network approach to building an affective module for a driver behavioural model}

\author[1]{Dorota Młynarczyk}
\author[1]{Gabriel Calvo}
\author[1]{Francisco Palmi-Perales}
\author[1]{Carmen Armero}
\author[2]{Virgilio Gómez-Rubio}
\author[3]{Ana de la Torre-García}
\author[3]{Ricardo Bayona Salvador}
\affil[1]{Departament d'Estadística i Investigació Operativa, Universitat de València, València, Spain}
\affil[2]{Departamento de Matemáticas, Universidad de Castilla-La Mancha, Albacete, Spain}
\affil[3]{Instituto de Biomecánica de València, Universitat Politècnica de València, València, Spain}

\date{}

\begin{document}

\maketitle
\begin{abstract}
This paper focuses on the affective component of a driver behavioural model (DBM). This component specifically models some drivers' mental states such as mental load and active fatigue, which may affect driving performance. We have used Bayesian networks (BNs) to explore the dependencies between various relevant random variables and assess the probability that a driver is in a particular mental state based on their physiological and demographic conditions. Through this approach, our goal is to improve our understanding of driver behaviour in dynamic environments, with potential applications in traffic safety and autonomous vehicle technologies.
\end{abstract}

\section*{Keywords}
Directed acyclic graph; Mental states; Probability and uncertainty; Statistical modelling.

\section{Introduction}
Understanding driver behaviour is a cornerstone of improving road safety, optimising traffic management, and developing autonomous vehicle systems. Human drivers exhibit complex decision-making processes influenced by numerous factors with a high degree of uncertainty, including environmental conditions, traffic scenarios, and psychological states among others. To reflect human behaviour, a probabilistic model is required that can both account for its complexity and capture the uncertainty associated with it. Bayesian networks (BNs) provide a robust framework to address this challenge.

A BN \cite{Pearl1995} is a probabilistic model within a graphical framework to represent  stochastic relationships between random variables. These variables are shown as nodes, and the edges between them encode probabilistic conditional dependencies, allowing the BN to assess probabilities associated with different outcomes based on data. From a modelling perspective, BNs provide a flexible framework for inference and prediction, because as \cite{pearl1995bayesian} state ``they are direct representations of the world, not of reasoning processes''. By incorporating probabilistic reasoning, BNs are   powerful tools for modelling complex systems. This approach is particularly suitable for analysing driver behaviour, as it can effectively accommodate the unpredictability inherent in human decision-making processes \cite{forbes1995batmobile}.

A Driver behavioural model (DBM) \cite{Negash2023} is a broad framework that integrates data from various sources, such as sensors, traffic systems, or driver profiles, to examine how different factors (such as speed, road conditions, and distractions) influence driver actions. This research is a part of the BERTHA EU project (BEhavioural ReplicaTion of Human drivers  \cite{bertha}), which aims to develop a comprehensive modular DBM capable of explaining and predicting human driving behaviour in a wide range of contexts relevant  for connected, cooperative, and automated mobility (CCAM). The DBM proposed in BERTHA integrates multiple dimensions of human driving—including cognitive, affective, perceptual, and contextual factors—within a unified probabilistic framework. By combining empirical data from controlled experiments and real-world driving, BERTHA seeks to capture the inherent variability of human driving.

In this paper, we focus on a part of the affective module of the DBM of the BERTHA project,   designed to capture and model the emotional and cognitive states that influence 
  the behaviour of a driver. Specifically, we will discuss   the intensity of some mental states, such as active fatigue and mental load, that significantly   impact a driver's response in various scenarios.  Furthermore, the model in this paper could easily be extended to explore how mental states influence driving actions, offering a more comprehensive understanding of driver behaviour under dynamic and uncertain conditions.

The remainder of this paper is organised as follows. Section~\ref{sec:BN} outlines the foundational principles of BNs in a Bayesian inferential context, details the database for the BN model and presents its structure. Section~\ref{sec:BBN} presents the results in general terms, illustrating them with predictive assessments of active fatigue and mental load. Finally, Section~\ref{sec:discussion} briefly discusses the extension of the model and areas for further research.

\section{Methods}
\label{sec:BN}
\subsection{Bayesian networks and Bayesian inference}

\begin{figure}[htbp]
\centerline{\includegraphics[width=4cm]{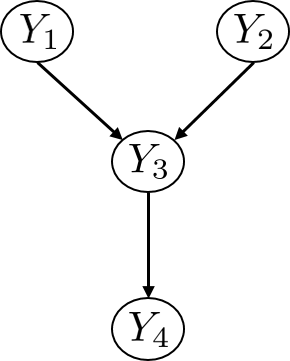}}
\caption{Basic direct acyclic graph example.}\label{figure_1}
\end{figure}

BNs are probabilistic models whose stochastic behaviour can be
depicted graphically by using directed acyclic graphs (DAGs).  
Graphs are mathematical structures that can be used to connect pairs of nodes by edges.  These elements are random variables and probability distributions in probabilistic DAGs, respectively. In addition, DAGs must adhere to two general properties: all edges must have a probabilistic direction, and the graph must be acyclic. In a DAG, the parents of a node are all those nodes in the network structure   that have directed edges pointing to it.  
Figure \ref{figure_1} illustrates the DAG corresponding to the set of random variables $\{Y_1, Y_2, Y_3, Y_4\}$, where the random variables  $Y_1$ and $Y_2$ are the parents of $Y_3$ (pa$(Y_3)=\{Y_1, Y_2\})$ and $Y_3$ is the only parent of $Y_4$ (pa$(Y_4)=\{Y_3\}$). The four variables are connected through three pointed edges that represent the probability distribution of $Y_3$ given its parents and that of $Y_4$ given $Y_3$. Nodes $Y_1$ and $Y_2$ have no parents.

Probabilistic models that depend on unknown parameters $\boldsymbol \theta$ need the statistic framework to estimate them.  In this paper,   this will be the Bayesian inferential methodology.  
  This approach requires the full specification of a probability distribution  for all random variables in the model, together with a prior distribution  for all the unknown parameters.
Formally, a BN \cite{pearl1995bayesian} is a joint probability distribution $f(\boldsymbol y, \boldsymbol \theta)$ over a set of random variables $\boldsymbol{Y} = \{Y_v : v=1, \ldots, V \}$,  representing the node set of the network, and a vector $\boldsymbol \theta$ of their subsequent parameters as follows:
\begin{align*}
 f(\boldsymbol{y}, \boldsymbol \theta) =  
 f( \boldsymbol y \mid \boldsymbol \theta)\, \pi(\boldsymbol \theta),
\end{align*}
\noindent where $f( \boldsymbol y \mid \boldsymbol \theta)$ is the conditional distribution of $\boldsymbol Y$ given $\boldsymbol \theta$ and $\pi(\boldsymbol \theta)$ is a prior distribution for $\boldsymbol \theta$.
 The BN satisfies the  factorization property. This means that   the conditional joint probability distribution $f(\boldsymbol y \mid \boldsymbol \theta)$ decomposes in terms of the univariate conditional distributions 
\begin{align*}
 f(\boldsymbol{y} \mid \boldsymbol \theta) = \prod_{v=1}^{V} f(y_v \mid \boldsymbol y_{\operatorname{pa}(Y_v)}, \boldsymbol \theta),
\end{align*}
\noindent being $f(y_v \mid \boldsymbol y_{\operatorname{pa}(Y_v)}, \boldsymbol \theta)$ the conditional distribution of node $Y_v$ given its parents and $\boldsymbol \theta$. Data $\mathcal D$ are the food of the statistical model that updates the information about $\boldsymbol \theta$ through Bayes' theorem to obtain its posterior distribution $\pi(\boldsymbol \theta \mid \mathcal D)$. The computation of this  posterior distribution is approximated via Markov chain Monte Carlo (MCMC)  simulation \cite{geyer2011}. This results in a subsequent approximate random sample that allows
  to approximate   the joint posterior predictive distribution 
$$f(\boldsymbol{y} \mid \boldsymbol{\mathcal{D}}) = \int \,f(\boldsymbol{y} \mid \boldsymbol{\theta}) \pi(\boldsymbol{\theta} \mid \mathcal{D})\,\mbox{d}\boldsymbol{\theta},$$
from which any posterior predictive marginal or conditional distribution   of interest can be approximated.

\subsection{Data structure and variables}

Data for the construction of a BN for the affective module of the DBM came from Instituto de Biomecánica de València (IBV),  the leading institution of the European BERTHA project. IBV performed an experimental study employing a driving simulator to characterise key mental states in car drivers. Mental workload and active fatigue were evaluated through a structured protocol combining subjective assessment with multimodal physiological monitoring. Participants completed validated mental workload and active fatigue scales at baseline and after each cognitive block, while physiological activity was recorded using ECG and respiratory signals via the PLUX Biosignals platform.

The resulting dataset encompasses a collection of demographic and physiological variables, together with responses to experimental questionnaires specifically designed to quantify mental load and active fatigue. The dataset comprises  observations collected from 17 participants, with the following basic covariate information:
\begin{itemize}
 \item $Y_{Sex}$: Sex;  8 women ($Y_{Sex}=0$) and 9 men ($Y_{Sex}=1$). 
 \item $Y_{Age}$: Age, in years; ranging from 24 to 61 (mean: 44.59 years).
 \item $Y_{BMI}$: Body mass index; ranging from 17.65 to 35.16 (mean: 24.35).
\end{itemize}

\noindent Physiological information   was assessed through  variables related to heart  and   respiration function:

\begin{itemize} 
 \item $Y_{SRT}$:   Standard Deviation of the Time elapsed between two successive normal heartbeats (RR intervals) in milliseconds (ms); ranging from 19.45 to 191.76  (mean: 63.07 ms).
 \item $Y_{SDD}$:  Standard Deviation of the successive Differences between adjacent RR intervals in milliseconds (ms); ranging from 14.38 to 104.16  (mean: 53.17 ms).
 \item $Y_{MHR}$:  Mean Heart Rate in beats per minute (bpm); ranging from 46.08 to 152.09 (mean: 73.94 bpm).
 \item $Y_{RLH}$: Ratio of the variance of the heart rate variability (HRV) in the Low-frequency to the High-frequency band; ranging from 0.17 to 29.91 (mean: 3.51).
 \item $Y_{MNB}$:  Mean of the Number of Breaths per minute; ranging from 9.42 to 23.38 (mean: 15.26 breaths/min).
\end{itemize}

Following the experimental tasks, participants also completed a series of  questionnaires regarding mental load, ($Y_{ML}$), and active fatigue, ($Y_{AF}$). Original responses were  transformed into a binary format with two  outcomes: 1, indicating that the participant entered the specified mental state, and 0 otherwise.

\subsection{A Bayesian network for assessing active fatigue and mental load in drivers}

\begin{figure}[htbp]
\centerline{\includegraphics[width=10cm]{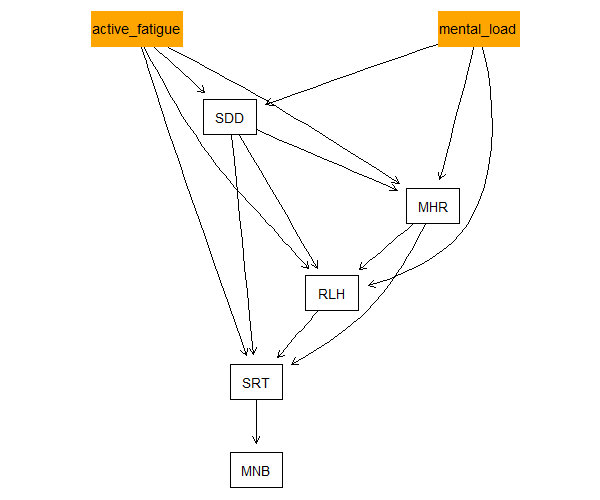}}
\caption{Bayesian Network for examining  active fatigue and mental load. For clearer visualization, the nodes have been identified  exclusively by the acronyms of their subindices.}\label{figure_2}
\end{figure}

The model was formulated as a BN  representing the probabilistic relationships among mental states, physiological responses,   covariates and parameters. Figure \ref{figure_2} presents the DAG of the analysed BN. It illustrates the relationships among mental load ($Y_{ML}$), active fatigue ($Y_{AF}$), and the set of physiological variables: $Y_{SRT}$, $Y_{SDD}$, $Y_{MHR}$, $Y_{RLH}$, and $Y_{MNB}$. For simplicity,  covariates and parameters have been omitted    from the graph. The nodes  corresponding to the two variables of interest, $Y_{ML}$  and $Y_{AF}$, have a direct influence on $Y_{MHR}$, $Y_{SDD}$, and $Y_{RLH}$. The node $Y_{SDD}$  affects $Y_{MHR}$, and both $Y_{SDD}$   and $Y_{MHR}$  contribute to $Y_{RLH}$. These three variables $Y_{MHR}$, $Y_{SDD}$, and $Y_{RLH}$ are parents of $Y_{SRT}$, which in turn is a parent of $Y_{MNB}$. These interdependencies illustrate the complex relationships within the network. In terms of interpreting these variables, it can be stated that, according to the model, mental states influence heart function, which in turn affects respiration. 
The subsequent conditional joint distribution of all nodes in the network given the set $\boldsymbol \theta$ of parameters factorizes as the product
\begin{align}
f&( y_{MNB},  y_{SRT}, y_{RLH}, y_{MHR}, y_{SDD}, y_{ML}, y_{AF}  \mid \boldsymbol \theta)= f( y_{MNB} \mid  y_{SRT},  \boldsymbol \theta) \nonumber\\
& \cdot f( y_{SRT} \mid y_{RLH}, y_{MHR}, y_{SDD}, \boldsymbol \theta) \cdot f(y_{RLH} \mid y_{MHR}, y_{SDD}, y_{AF}, y_{ML},\boldsymbol \theta)  \nonumber\\
& \cdot f(y_{MHR}\mid   y_{SDD}, y_{AF}, y_{ML},\boldsymbol \theta) \cdot f( y_{SDD} \mid     y_{AF}, y_{ML},\boldsymbol \theta) \nonumber \\ & \cdot f(  y_{AF} \mid \boldsymbol \theta) \cdot f( y_{ML} \mid \boldsymbol \theta).  
\label{eqn:producto}
\end{align}

This decomposition allows each  univariate conditional distribution in (\ref{eqn:producto}) to be modelled independently. Mental load ($Y_{ML}$) and active fatigue ($Y_{AF}$)  are Bernoulli variables whose probability of success is modelled using Bayesian logistic regression (BLR). We analysed the conditional distribution of each of the continuous physiological measures, $Y_{MNB}$, $Y_{SRT}$, $Y_{RLH}$, $Y_{MHR}$ and $Y_{SDD}$ in (1) using a Bayesian   regression model (BRM) with the subsequent conditional mean expressed as a linear function of the three base covariates and the parents variables of the response variable considered. For example, in the case of the modelling of $f(y_{MHR}\mid   y_{SDD}, y_{AF}, y_{ML},\boldsymbol \theta)$ we assume a normal distribution with mean $\mu_{MHR}$ expressed as    a linear function of the three base covariates, $Y_{Sex}$, $Y_{Age}$, and $Y_{BMI}$, as well as nodes  $Y_{SDD}$, $Y_{AF}$, $Y_{ML}$ as the parents of $Y_{MHR}$.
\begin{align}
    \mu_{MHR}   =  &  \beta_0+\beta_W I_W + \beta_{Age} Y_{Age}+ \beta_{BMI} Y_{BMI} +\nonumber \\
     &    \beta_{SDD} Y_{SDD} + \beta_{AF} Y_{AF} +  \beta_{ML} Y_{ML},   
\end{align}
where $I_W$ is an indicator variable with value 1 if the observation comes from a man and zero otherwise and $\beta$'s are the regression coefficients, parameters,  associated with the covariates.  

The prior distributions for the model parameters $\boldsymbol \theta$  assume  prior  independence in a non-informative environment that gives all inferential prominence to the data. All regression coefficients    $\beta$'s are assigned normal distributions with mean 0 and variance 25 in all BLR and BRM models.   The standard deviation parameters  of the BRM models are elicited as uniform distributions in the interval $(0,30)$.

\section{Results}
\label{sec:BBN}
The posterior distribution $\pi(\boldsymbol \theta \mid \mathcal D)$ for all parameters of the BN was approximated by MCMC through \texttt{JAGS} software. The subsequent posterior MCMC sample was used to approximate the joint posterior predictive distribution. This results in a sample from which we can extract the behaviour of any predictive element of interest.

\begin{figure}[htbp]
\centerline{\includegraphics[width=16cm]{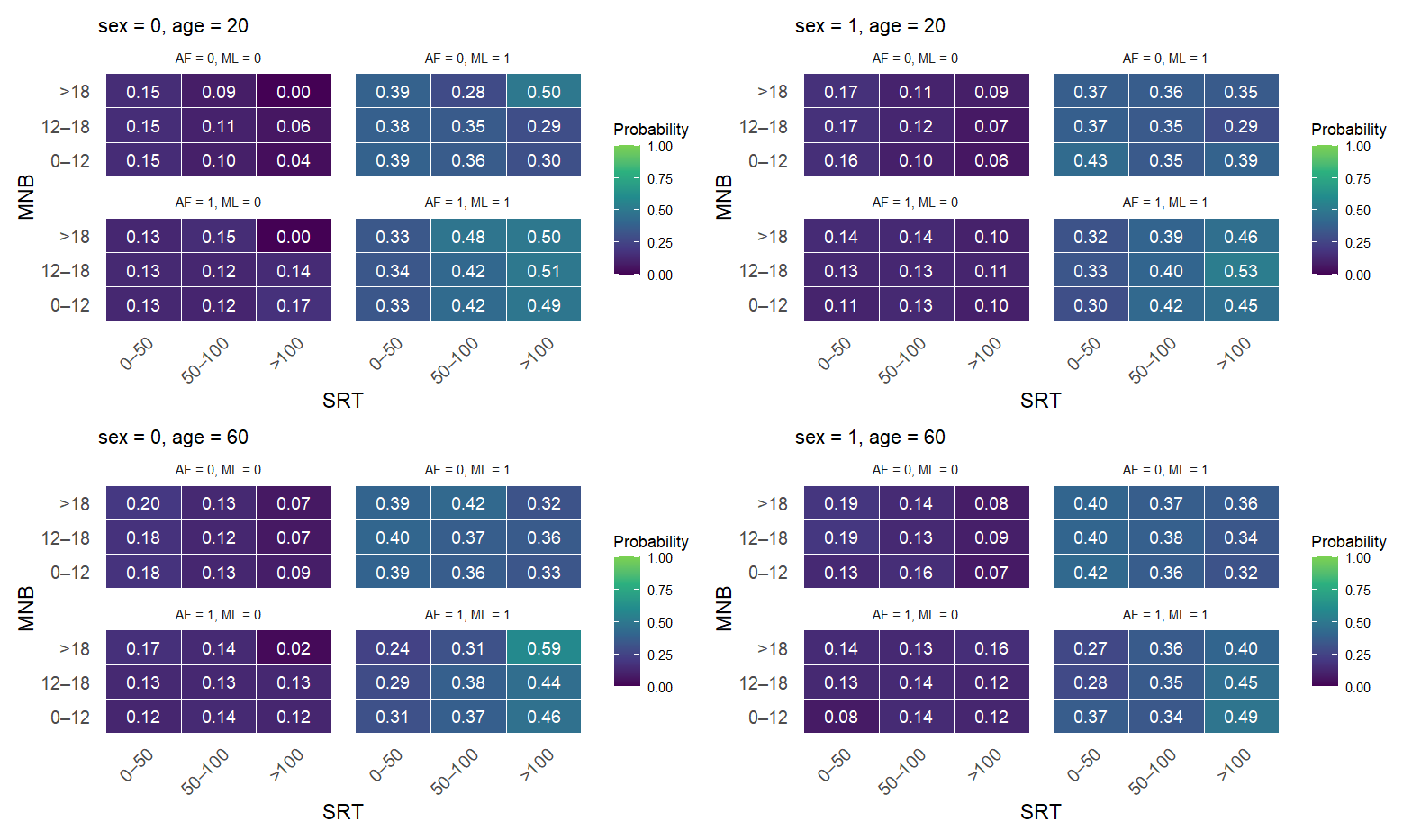}}
\caption{Average posterior probability of each joint state of active fatigue (AF) and mental workload (ML) as a function of driver gender and age, with BMI held constant at 22, shown across  $Y_{SRT}$ and $Y_{MNB}$ values.
}\label{figure_3}
\end{figure}

By way of illustration, Figure \ref{figure_3} shows the average posterior probability that a driver exhibits each of the four conditions defined by active fatigue and mental workload in relation to their gender, age, and BMI, and the range of possible values for the $Y_{SRT}$  and $Y_{MNB}$ nodes. In a real application, conditional posterior probabilities could be derived; for example, for a female driver aged 20 with a BMI of 22 and a mean breathing rate of 20 breaths per minute, the probability of experiencing mental workload is estimated to be approximately 0.72. Predictions seem to indicate that age has a protective effect against active fatigue and mental workload, but not gender, which does not appear to be relevant. 

\section{Discussion}
\label{sec:discussion}
The ability of BNs to estimate probabilities associated with specific mental states, such as mental load or active fatigue, under different physiological conditions of the driver is an important aspect of this study. This information would enable real-time assessments of the driver's mental state, with important implications for road safety \cite{liu2020analysis}. For example, a vehicle could suggest a break to a driver showing signs of mental fatigue. This model could help optimize the design of systems to prevent accidents caused by cognitive fatigue or mental overload.

The results of this study show promising results, but the complexity of the subject requires further research and more data. They   were obtained from a fairly small sample, and it would be desirable to work with larger samples that could provide greater power and predictive efficacy. It could be interesting to explore additional markers, such as facial expressions, which could provide additional information about the driver's mental state. 

\section*{Acknowledgments}
Research activity under BERTHA project (GA101076360) funded by the European Union. Views and opinions expressed are however those of the author(s) only and do not necessarily reflect those of the European Union or the European Climate, Infrastructure and Environment Executive Agency (CINEA). Neither the European Union nor the granting authority can be held responsible for them.

\end{document}